\documentclass[twocolumn]{aastex631}
\usepackage{xcolor}
\usepackage{natbib}
\usepackage{hyperref}
\usepackage{amsmath,mathtools}
\usepackage{comment}

\hypersetup{
    pdfnewwindow=true,      % links in new window
    colorlinks=true,        % false: boxed links; true: colored links
    linkcolor=blue,         % color of internal links
    citecolor=black,        % color of links to bibliography
    filecolor=blue,         % color of file links
    urlcolor=blue           % color of external links
}

\newcommand{\Mo}{M_{\odot}}

\begin{document}
\def\gsim{ \lower .75ex \hbox{$\sim$} \llap{\raise .27ex \hbox{$>$}} }
\def\lsim{ \lower .75ex\hbox{$\sim$} \llap{\raise .27ex \hbox{$<$}} }
\def\clov#1{{\bf[#1 -- Clovis]}}

\shorttitle{Low and High Energy Neutrinos from SN 2023ixf}

%\shorttitle{High Energy neutrino emission from chocked jet in Supernovae explosions}
\shortauthors{Guetta, Langella, Gagliardini, Della Valle}
 
%\title{High-energy Neutrino Emission from Chocked Jets in Core-collapse Supernovae}
\title{Low and High Energy Neutrinos from SN 2023ixf in M101}

%\correspondingauthor{Dafne Guetta}

\author{Dafne Guetta}
\email{dafneguetta@gmail.com}
\affiliation{Physics Department, Ariel University, Ariel, Israel}
\author{Aurora Langella}
\email{alangella@na.infn.it}
\affiliation{University of Naples Federico II, Napoli, Italy and INFN Naples}
\author{Silvia Gagliardini}
\email{
silvia.gagliardini@roma1.infn.it}
\affiliation{Physics Department, Ariel University, Ariel, Israel}
\affiliation{Istituto Nazionale di Fisica Nucleare, Sezione di Roma, P. le Aldo Moro 2, I-00185 Rome, Italy}
\author{Massimo Della Valle}
\email{massimo.dellavalle@inaf.it}
\affiliation{Capodimonte Observatory, INAF-Naples , Salita Moiariello 16, 80131-Naples, Italy}
\affiliation{Physics Department, Ariel University, Ariel, Israel}
%\author{Gianfranca Derosa}
%\email{gderosa@na.infn.it}
%\affiliation{University of Naples Federico II, %Napoli, Italy and INFN Naples}

\begin{abstract}
%Long duration Gamma-ray Bursts are associated with massive stars that have lost their H and He envelopes when the gravitational collapse of the core occurs, and then they explode as Supernovae Ic (SNe-Ic). During the collapse of the core,  collimated outflows (jets) along the rotational axis  are formed. 

Supernova (SN) 2023ixf in M101 is the closest SN explosion observed in the last decade. Therefore it is a suitable test bed to study the role of jets in powering the SN ejecta. With this aim, we explored the idea that high-energy neutrinos could be produced during the interaction between the jets and the intense radiation field produced in the SN explosion and eventually be observed by the IceCube neutrino telescope. The lack of detection of such neutrinos has significantly constrained both the fraction of stellar collapses that produce jets and/or the theoretical models for neutrino production. Finally, we investigated the possibility of detecting low-energy neutrinos from SN 2023ixf with the Super- and Hyper-Kamiokande experiments, obtaining in both cases sub-threshold estimates.
\end{abstract}

\keywords{gamma rays: bursts --- stars: binaries --- stars: neutron --- gravitational waves}

\section{Introduction}
When in a massive star (at least M $> 8M_{\odot}$) the gravitational collapse of the core occurs, while the star still retains its H envelope, we witness a type II supernova explosion. Almost all of the gravitational binding energy, about $ \sim 10^{53}$ erg, released during the formation of a neutron star is radiated as neutrinos. The kinetic energy associated with the moving ejecta, at $\sim 10,000$ km/s, accounts only $\sim 10^{51}$ erg of the entire energy budget, while $\sim 10^{49}$ erg goes into the "luminous energy" (lightcurve) observable with our telescopes. The commonly accepted mechanism to explain the SN explosion is the so-called "neutrino re-heating", discussed since the '60s by several authors \citep{colgatewhite,colgatemkee,Colgate82,Bahcall1990,janka1996}. After the inner part of the core has collapsed and formed the proto-neutron star causing the infalling material to bounce, an outward propagating shock is formed and would soon stall, without "neutrino heating". Indeed, if only a tiny fraction of the neutrinos energy is reabsorbed by surrounding material, the star expansion could resume. This expansion can trigger a violent shock wave that moves through the progenitor star, then causing the supernova explosion. The neutronization process that takes place after the collapse of the core, received a splendid confirmation from the observation of 25 neutrinos associated with SN 1987A \citep{Hirata87,Bionta87,Alexeyev88}
out of $\sim 10^{58}$ emitted neutrinos of all flavors with typical energies of $\sim 10$ MeV.

In recent years, another mechanism actually hypothesized a few decades ago \citep{Bisnovatyi1970,LeBlanc1970,Ostriker1971}, but which is perhaps experiencing a second youth thanks to studies on long-duration GRBs, is gaining consensus within the scientific community. The idea is that jets could play an important role not only in the explosion of massive long-duration GRB progenitors but also in other core-collapse supernovae (CC-SNe) \citep{Piran2019,Soker22}. For the sake of clarity, we briefly summarize the ideas underlying this scenario.  It is possible that during all core-collapse events, jets might be produced and launched immediately after the collapse of the core. If this occurs when the progenitor has lost its H and He envelopes, we may observe a type Ic SN event associated with a long-duration GRB \citep{Meszaros2001Jet,Woosley2006,Hjorth011}. When the collapse of the core takes place while the progenitor is still retaining the H/He envelopes the jets are unable to emerge from the star due to the presence of these massive envelopes. These "chocked" jets remain trapped within the progenitor and  during their interaction with the stellar material, create a "cocoon" where the jet deposits up to $\sim 10^{51-52}$ erg \citep{Nakar2017}. These energies are comparable to the kinetic energies which characterize the ejecta of "standard" CC-SNe and GRB-SNe respectively. An important piece of evidence in favor of this alternative scenario is that in 2019 the spectroscopic signatures of a jet cocoon were actually observed for the first time in the core-collapse SN 2017iuk \citep{Izzo2019}. 
Due to inhomogeneities, internal shocks may occur inside the jet at a radius $R$ smaller than the jet head radius $ R_{\rm H}$. The protons present in the jet can be accelerated in the internal  shock region and interact with the thermal photons present in the stellar envelope leading to the production of charged pions that can decay into high-energy neutrinos \citep{Meszaros2001Jet}. If this jet is not energetic enough to break through the remaining outer layers of the star and stalls sufficiently far below the photosphere, the high energy neutrinos produced by accelerated protons would be the only messengers able to escape \citep{Senno2016,Murase2016,He2018,Guetta2020,Fasano2021} and in principle could be observed by IceCube. 
The contribution of CC-SNe to the diffuse neutrino flux detected by IceCube has recently been explored. \cite{icecube_2023}. From the stacking analysis performed using the data recorded by the IceCube Neutrino Observatory no significant temporal and spatial correlation of neutrinos and SNe was found \cite{icecube_2023}. The search for high-energy neutrinos from SN PS16cgx was equally unsuccessful (\cite{kankare2019}.

Another model for high-energy neutrino production that has been considered in this paper, is the one described in \cite{Murase2011} (see also \citep{Murase2018,Murase2022,Murase2023,valtonen2023} for updates to this model.) They consider the collisions of the SN ejecta with the massive circumstellar medium (CSM) shells as potential proton accelerators. The accelerated protons may interact with the protons present in the dense CSM and produce high-energy neutrinos. If the energy in accelerated protons is only $\sim 10\%$ of the SN energy,  multi-TeV neutrinos should be detected at later epochs than the optical/infrared luminosity peaks.

In this paper, we consider the recent case of SN2023 ixf, which occurred in the nearby galaxy M101, as a test bed for possible detection of neutrinos by CC-SN events. SN 2023ixf was discovered by K. Itagaki in M101 (d=6.4 Mpc; \citep{Shappee2011}) on May 19, 2023, and has been classified as a Type II SN by \citep{Perley2023}. A preliminary discussion of the properties of this SN has been presented by \cite{Yamanaka2023}. These authors show a series of spectra obtained at very early stages (t$\sim 2$d) characterized by the presence of narrow line components (e.g. H$_\alpha$, HeII, NIV, and CIV) which are signatures of the interactions of the SN ejecta with the dense CSM \citep{Neustadt2023,Cappellaro2015,Jacobson-Galan2023} . However, these narrow components have quickly vanished (see for example Fig. 4 of \cite{Yamanaka2023}) likely within two weeks from maximum \citep{Bostroem2023}. This piece of observation  indicates that the circumburst material was very close to the progenitor and the interaction with the SN ejecta practically occurred in the aftermath of the SN explosion, significantly limiting the temporal window for the detection of "delayed neutrinos". In this work, we estimate the number of neutrino events expected from this SN both at low energy ($\sim$ MeV) from Super-kamiokande (Super-K) and Hyper-kamiokande (Hyper-K) (§2) and at high energy from IceCube (§3). In §4 we discuss the observational predictions. In section §5 we report our Conclusions. 

\section{Low energy neutrinos}

CC-SNe are one of the most powerful cosmic sources of neutrinos in the Universe. It is expected that, during the burst,  99\% of their energy is emitted as neutrinos with energies in the MeV energy band.
All flavors of neutrinos and anti-neutrinos are expected to be produced. The most favored model for the CC explosion mechanism is the “delayed neutrino-heating mechanism” discussed in \cite{Bethe:1985sox}.
At the beginning of the stellar collapse, a few milliseconds after the core bounce, a prompt burst of electron neutrinos is emitted, which, alongside iron photo-disintegration, drains energy from the post-shock material and reduces the pressure behind the shock. Thus, the expanding shock is replaced by an accretion shock characterized by matter in-falling. 
%This lead to the failure of the so called "prompt bounce-shock mechanism" (for a review, see \cite{art:Janka2012}).
While the core density increases (until hundreds of milliseconds after the bounce), neutrinos get trapped in the core, transferring energy to the surrounding material. This process is known as "neutrino re-heating". Neutrino-energy deposition behind the stalled bounce shock can revive the shock and thus initiate the SN explosion. Although in the last decades, many constraints on neutrino physics and supernovae could be derived on the basis of the SN 1987A neutrino observation, the limited statistics of only two dozen registered events prevented a full understanding of the mechanism that drives a supernova explosion. Many models have been proposed to simulate the expected neutrino emission from supernovae, employing a variety of different approximations and simplifying assumptions. This leads to different outcomes when simulating the supernova explosion and the consequent number of expected neutrinos. Thus, the next supernova neutrino detection will be crucial to test the different supernova models and to better understand the explosion mechanism. In this section, we estimate the number of neutrinos expected from SN events that can be detected by Super-K and Hyper-K, as a function of their distance (Fig. 1). 

\subsection{Super-Kamiokande}
Super-Kamiokade is a water Cherenkov detector located in Gifu Prefecture, Japan \citep{art:FUKUDA2003}. 
It consists of a cylindrical stainless-steel tank of 39m diameter and 42m height, filled with 50 kton of water and with about 13k photo-multiplier tubes (PMTs) covering its surfaces, in order to detect charged particles produced by neutrino interactions via Cherenkov radiation. Thanks to its huge volume and low threshold (few MeVs), it is one of the neutrino experiments with the highest sensitivity to supernova neutrinos. \\ In the energy range of interest for supernovae, up to $\sim100$ MeV,  one of the main sources of background noise comes from the radioactive impurities of the PMTs, the detector walls, and the surrounding rocks. Therefore, to increase the sensitivity, low-energy events with a reconstructed vertex within 2 m from the walls are discarded. This leads to an actual fiducial volume (FV) of 22.5 kton \citep{art:FUKUDA2003}.  In the following, we will always assume a volume of 22.5 kton when evaluating the expected number of neutrinos from SN2023ixf in Super-K. \\
In the low energy range, the main interactions in Super-K are: 
\begin{itemize}
    \item Inverse Beta Decay (IBD): 
    $p+\overline{\nu}_e\longrightarrow n+e^+$
    \item $\nu$e-scattering: $\nu+e\longrightarrow \nu+e$,
    \item  $\nu_e \prescript{16}{}{O}$ Charge-Current (CC) interaction: 
    $\nu_e + \prescript{16}{}{O}\longrightarrow e^- + \prescript{16}{}{F}^{(*)}$
    \item $\overline{\nu}_e \prescript{16}{}{O}$ CC interaction: 
    $\overline{\nu}_e + \prescript{16}{}{O}\longrightarrow e^+ + \prescript{16}{}{N}^{(*)}$
    
\end{itemize}

\begin{figure}
    \centering
    \includegraphics[scale=0.55]{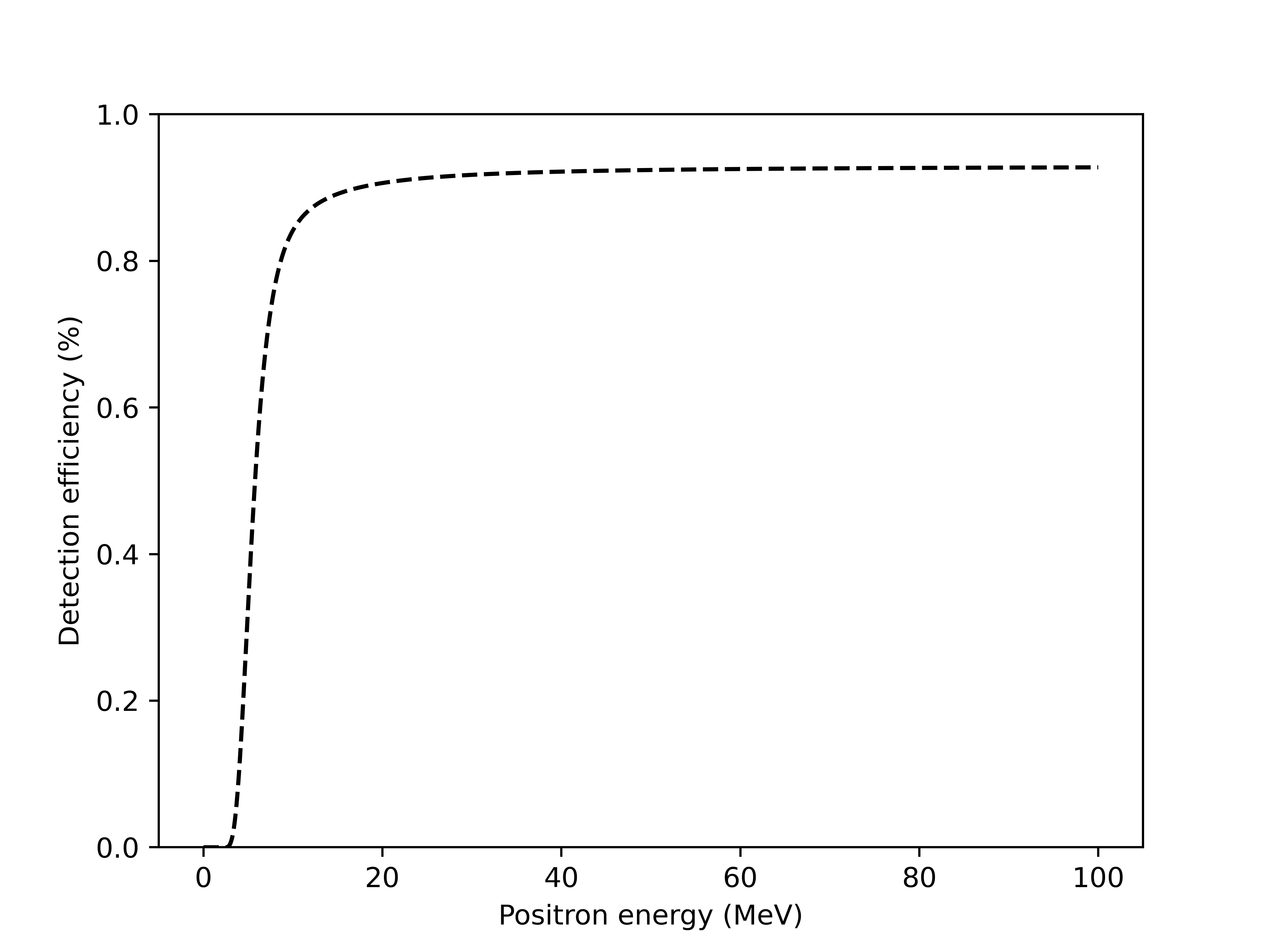}
    \caption{Super-Kamiokande detection efficiency as a function of the positron energy. Taken from \cite{art:vissani2022}.}
    \label{fig:eff}
\end{figure}

Since the dominant reaction (about 90\% of all events) of supernova neutrinos with water is IBD, we considered only events produced by IBD in this work.\\ In this work we have considered the Super-K detection efficiency derived from Eq. 4.4 of  \cite{art:vissani2022} (considering a threshold of 5 MeV). We show this efficiency as a function of energy  in Figure \ref{fig:eff}.\\

We adopted four different models to evaluate the expected number of events in Super-K from SN2023ixf.\\

The Livermore model (or Totani model) \citep{art:Totani_1998} uses a 20 $M_{\odot}$ progenitor, which was modeled to reproduce the light curve of SN1987A, and a simulation code developed by Wilson and Mayle \citep{art:Wilson1986,art:Mayle1987}. The simulation is one-dimensional, assumes an amount of energy radiated in electron antineutrinos equal to $L_{\bar{\nu}_e}=2.9\cdot 10^{53} erg$ and is performed from the start of the collapse to 18 s after the core bounce. Although outdated, it represents a baseline supernova model as it is the first simulation based on SN1987A data and it is one of the few SN models that provide predictions of neutrino emission until supernova late stages. \\

The Nakazato model \citep{Nakazato_2013} is a family of models that includes different progenitor masses (13-50 $\Mo$) and metallicities ($Z=0.02, 0.004$). It has been widely used in many supernova studies, e.g. \cite{DSNB, Jia_2019, Li2019, Kistler2013}. Similar to the Livermore model, it performs a 1-D simulation and can describe neutrino emission until 20 s after the bounce but it implements a more advanced description of neutrino transport and can take into account neutrino oscillation during the propagation. \\
Based on \cite{Kilpatrick2023}, the progenitor of SN2023ixf is consistent with RSG models of initial mass 11 $\Mo$, therefore we considered in the simulation a progenitor mass of $13 \Mo$, with a metallicity of $Z=0.02$ and a shock revival time of 300 ms. The total energy emitted by electron anti-neutrinos is $L_{\bar{\nu}_e}=7\cdot10^{52} erg$. Given the distance of SN2023ixf, neutrino oscillation doesn't impact significantly the expected number of anti-electron neutrinos at Earth, thus we didn't consider this effect in our scenario.\\

The Fornax (2021) model \citep{Fornax_2021} is a 2-D simulation that considers progenitors with solar metallicity (Z=0.02) and masses in the range [12,26]$ \Mo$. It simulates neutrino emission until 4.5 s post-bounce using the FORNAX code \citep{Skinner_2019}. With respect to 1-D models, 2-D simulations allow to describe the turbulent convection in the core while requiring less computational time than 3-D simulations. We adopted the same initial mass as in the Nakazato model, i.e. 13 $\Mo$ while the total energy emitted  by electron anti-neutrinos in this model is around $8\cdot 10^{52}$ erg. As before, neutrino oscillation was omitted.\\

We refer to \cite{art:ikeda} to evaluate the number of neutrinos expected from SN2023ixf based on the Livermore model given its distance ($D=6.4$ Mpc) and a threshold of 5 MeV. \\For Nakazato and Fornax (2021) models,  we adopted the public tool SNEWPY \citep{Baxter2021}. It consists of a Python-based code that allows to study the detection prospects of supernova neutrinos in many neutrino detectors, based on different supernova models. In this pipeline, the detectors' response is simulated using  the SNOwGLOBES tool \citep{snowglobes}.\\
 
\begin{figure}
    \centering
    \includegraphics[scale=0.55]{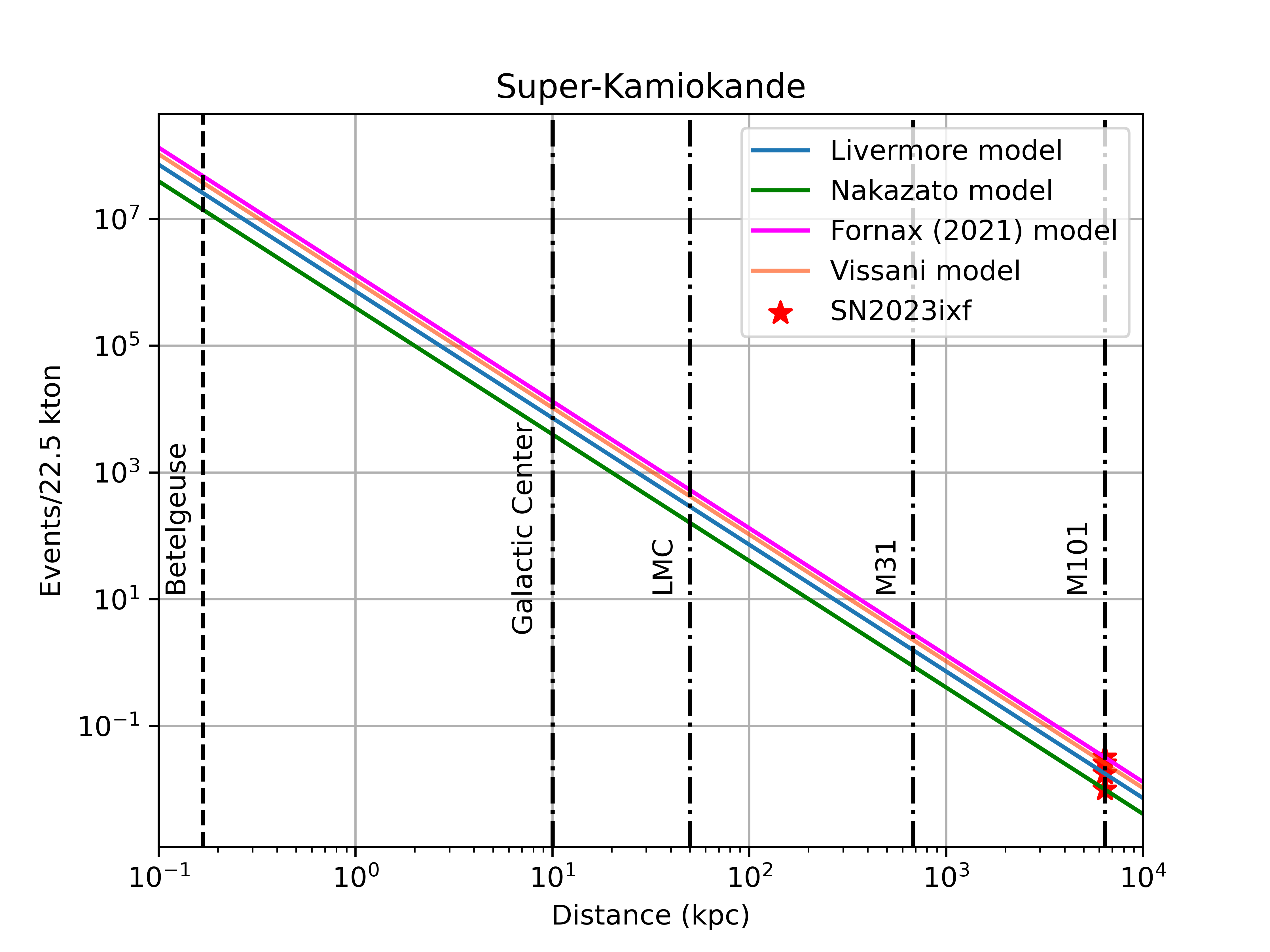}
    \caption{Number of expected IBD events in Super-Kamiokande as a function of distance for supernovae based on different models. The number of events is calculated considering the fiducial volume (FV) of the detector (22.5 kton). Black dashed lines correspond to "typical" distances within the Milky Way and in the Local Group, up to to M101, the host galaxy of SN2023ixf.}
    \label{fig:sk_events}
\end{figure}

Last model adopted (indicated in the following as Vissani model) is based on \cite{art:Vissani2015}. 
In this paper, rather than numerically simulate the explosion of a supernova, they propose a parameterized expression for the electron anti-neutrino flux expected at Earth, based on SN1987A data.\\ 
By following the methodology described in Section 2 of \cite{art:Vissani2015} while considering the updated IBD cross section included in \cite{art:vissani2022}, we assumed $L_{\bar{\nu}_e}=10^{53}$  and $D=6.4$ Mpc and calculated the expected IBD positron spectrum in Super-K derived from anti-electron neutrinos produced by SN2023ixf. Then, by varying the distance of the supernova and integrating the spectrum in the range [5, 100] MeV, the number of expected events as a function of distance for Super-K has been evaluated. \\ 
Figure \ref{fig:sk_events} shows the expected number of IBD events in Super-K as a function of the supernova distance for all the models tested. For comparison, two galaxies of the Local Group are indicated in the plot alongside the galaxy hosting SN2023ixf.\\
In general, we obtained a very low number of expected events from SN2023ixf, with values in the range $N_{\bar{\nu}_e}=[0.01, 0.03]$. \\
In this work, it has not been considered that Super-K has recently entered a new phase, named SK-Gd, where a percentage of Gadolinium (Gd) has been added to the pure water in the tank \citep{art:sk-gd}. The presence of Gd enhances the detector’s ability to identify and differentiate antineutrino-induced events from backgrounds from IBD reactions, thus possibly improving the sensitivity to SN neutrinos.
\smallskip
The non-detectability of this supernova by Super-K has been confirmed by the Super-K Collaboration in \cite{atel_sk}, as they performed a search for neutrino signal correlated to SN2023ixf in a time window of two days prior to the SN and no excess signal was found. 

\subsection{Hyper-Kamiokande}

Hyper-Kamiokande will be the next generation underwater Cherenkov detector, whose operations will start in 2027 \citep{art:HK}. While the baseline design of Hyper-K represents an improvement of the highly successful Super-K, it will be characterized by a volume that is around 8 times bigger and an improved photosensor configuration, allowing for a sensitivity that is far beyond that of its predecessor, especially for what concerns astrophysical neutrinos. 
 To evaluate the number of neutrinos expected from SN2023ixf in Hyper-K, we assumed the same detection efficiency as in Super-K (see Fig. \ref{fig:eff}) while considering an increased volume. \\
%and assumed a similar signal efficiency.}\\
Hyper-K's response to SN event based on the Livermore model has been extracted from \cite{art:HK} (where a 0.56 Mton fiducial volume is assumed) and re-scaled it considering the current expected fiducial volume in Hyper-K (0.22 Mton) reported in \cite{art:HK_sn}. \\
For the Nakazato, Fornax (2021), and Vissani models we used the same tools and methodology as in the previous paragraph but scaled the results considering Hyper-K's volume (0.22 Mton). \\
We also adopted a fifth model for Hyper-K, based on \cite{nakamura}. In this model, it is considered a non-rotating solar-metallicity progenitor with a mass of 17$\Mo$, which retains its hydrogen envelope (thus a type II supernova is expected). In this case, the simulation produces neutrino emission for the first 7 s after the core bounce. Neutrino MSW mixing is considered during propagation in the star's envelopes. Only IBD events in the range 18-30 MeV are considered for Hyper-K. \\Hyper-K's response to CCSNs in nearby galaxies (few Mpc) is approximated as follows\footnote{The formula is a corrected version of Eq.5 in \cite{nakamura}, obtained after a private communication with Ko Nakamura.}: 
\begin{equation}
    N_{\bar{\nu}_e}\simeq8.5\left(\dfrac{D}{1Mpc}\right)^{-2}\left(\dfrac{M_{det}}{0.56Mton}\right)
\end{equation}

\begin{figure}
    \centering
    \includegraphics[scale=0.55]{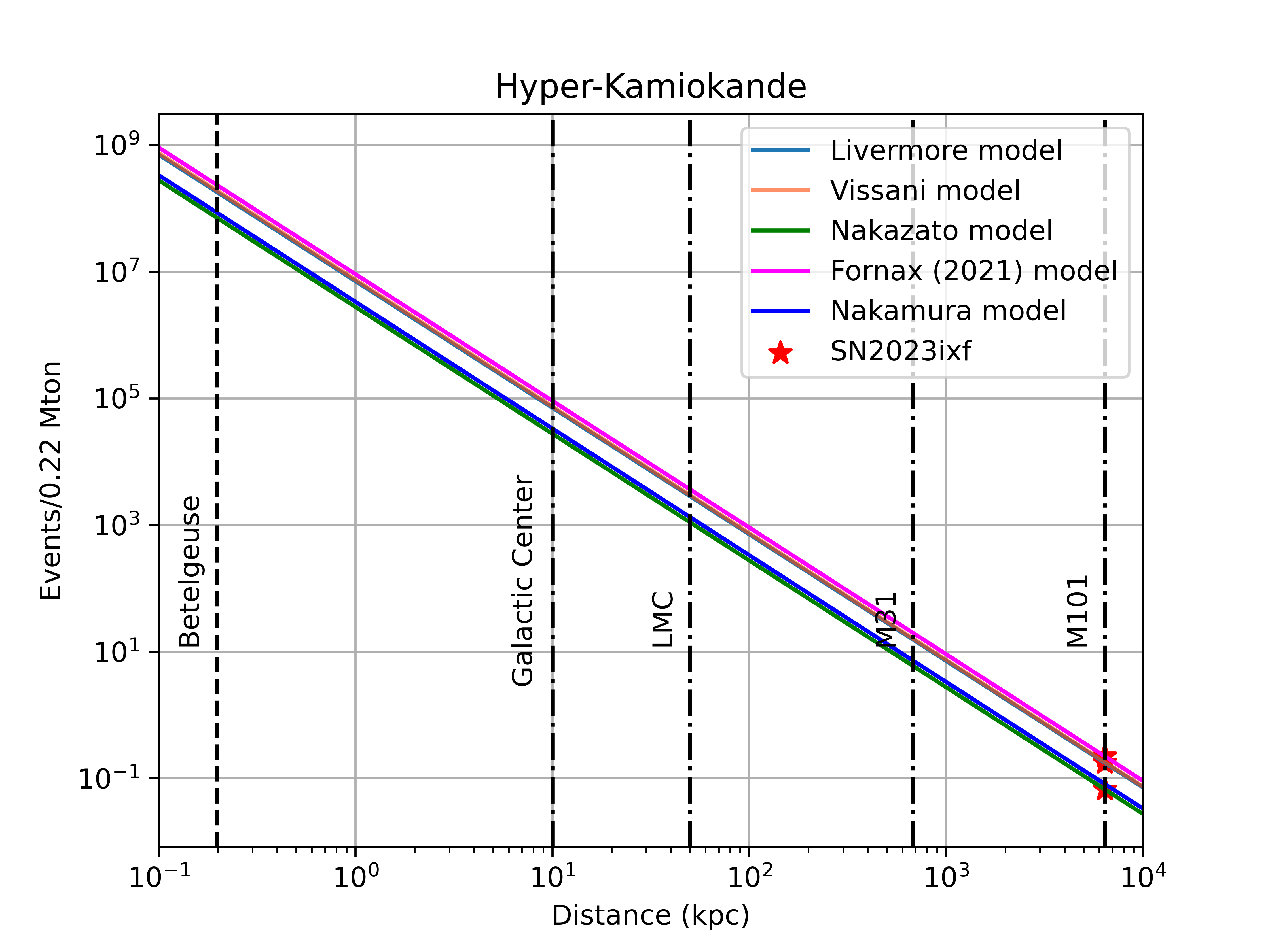}
    \caption{Number of expected IBD events in Hyper-Kamiokande as a function of distance for supernovae based on different models. The number of events is calculated considering the FV of the detector (220 kton). Black dashed lines correspond to "typical" distances within the Milky Way and in the Local Group, up to to M101, the host galaxy of SN2023ixf.}
    \label{fig:hk_events}
\end{figure}

Fig.\ref{fig:hk_events} shows the number of IBD events expected in Hyper-K as a function of supernova distance based only on all the models considered. We obtain  a similar number of events in all cases, with values in the range $N_{\bar{\nu}_e}=[0.07, 0.22]$. 

\section{High energy neutrinos from chocked jets}

We assume for SN 2023ixf a Red Supergiant (RSG) progenitor of size $\sim 3 \times 10^{13}$ cm \citep{Kilpatrick2023,Jencson2023,Hosseinzadeh2023} with a density $\rho_H\sim 10^{-7} {\rm g}{cm}^{-3} \rho_{H,-7}$, where $\rho_{H¸-7}=\left(\rho_H/10^{-7}\right)$\citep{Meszaros2001Jet,He2018,Guetta2020,Fasano2021}. After the collapse of the core, the newborn neutron star (NS) accretes mass from the innermost stellar material and launches energetic jets that cannot break through the progenitor star due to the H/He envelopes and remaining stuck inside, then forming the so-called "chocked" jets \citep{Piran2019,Meszaros2001Jet, Macfadyen2001}. A forward shock and a reverse shock (RS) are produced when the jet is propagating in the hydrogen envelope. In the region of the RS, electrons can be accelerated and emit photons that are free to move inside the jet \citep{Senno2016}, however, they cannot leave the source as the envelopes or circumstellar medium outside are largely optically thick to Thomson scattering. Due to inhomogeneities of the jet, internal shocks can be produced and  dissipate a fraction of the jet kinetic energy, $f_p$ to accelerate protons.
The protons accelerated at the internal shock region can interact with the thermal photons of the RS region very efficiently since the latter are trapped inside the jet. Therefore the photo-meson interaction is very efficient, implying that the fraction of protons converted in pions, is almost 100$\%$ in this process. The pions then decay into high-energy neutrinos. 

In this paper, we consider this "chocked jet" model \citep{Senno2016,He2018,Fasano2021} to estimate the neutrino flux and spectrum  from SN2023ixf. We take into account the high energy cutoff due to the photo-meson cooling of protons, and the synchrotron cooling of pions and muons.

Following \citep{He2018,Fasano2021}, we adopt a numerical simulation to calculate the neutrino spectrum distribution due to the photomeson interaction between accelerated protons and target photons. We generate a flux of accelerated protons, in the IS region, with energies from 100 GeV up to $10^{9}$~GeV, according to the spectrum described by Fermi I-order acceleration process as:

\begin{equation}
    \frac{dN_{\rm p}}{dE_{\rm p}} \propto E_{\rm p}^{-2}
\end{equation}

In the RS region a fraction of energy, $\epsilon_e\sim0.1$, is dissipated in accelerating electrons that lose their energy in synchrotron emission. Due to the large optical thickness, the photons generated from this process thermalize at $T_{\gamma}$:

\begin{equation}
    k T_{\gamma} \sim 313 eV L_{50}^{1/8} \epsilon_{e,-1}^{1/4} t_{jet,3}^{-1/4} \rho_{H,-7}^{1/8}
\end{equation}
where $L$ and $t_{\rm jet}$ are the jet luminosity and duration.
The target photon spectrum  in the RS region is well represented by a blackbody distribution:

\begin{equation}
    \frac{dN_{\gamma}}{dE_{\gamma}}= \frac{8}{19(k T_{\gamma})^3} \frac{E_{\gamma}^2}{e^{\frac{E_\gamma}{k T_\gamma}}-1}
\end{equation}

where the peak energy distribution is given by Wien's displacement law:

\begin{equation}
    E_{\gamma}^{max} \sim 2.82 k T_{\gamma} \sim 881 eV
\end{equation}

The photon density  follows the Planck distribution:

\begin{equation}
    n_{\gamma}= \int_0^\infty \frac{dN_{\gamma}}{dE_{\gamma}} dE_{\gamma} \sim 16 \pi \zeta(3) (\frac{kT_{\gamma}}{h c})^3
\end{equation}

\noindent

where $\zeta(3)$ is the Riemann zeta function.
%A fraction of thermal photons will then escape in the IS, since it is optically thin, where their peak energy becomes:

%\begin{equation}
    %E_{\gamma,IS}^{max}= \Gamma_{IR}E_{\gamma,RS}^{max} 
%\end{equation}

%where $\Gamma_{IR} \sim \Gamma$ is the Lorentz factor of the IS respect to RS.

The photon density in the internal shock (IS) frame can be obtained as \cite{Fasano2021}:

\begin{equation}
 n_{\gamma,\rm IS}= \Gamma n_{\gamma} f_{\rm esc} \sim 1.9 \times 10^{20} cm^{-3} \Gamma ^2_2 L^{-3/4}_{50} t^{1/2}_{\rm jet,3} \rho^{-1/4}_{H,-7}
\end{equation}

The parameters that affect the neutrino flux are the kinetic energy of the jet, $E$, the Lorentz factor, $\Gamma$, the duration of the jet, $t_{\rm jet}$ and consequently the luminosity of the jet, $L=E/t_{\rm jet}$.  
We consider $E=10^{51},10^{53}$ erg, and different time durations for the jet \citep{He2018} $t=100,10^3, 10^{4}$\,s. 
The velocity of the jet is an unknown parameter, it can be assumed to be  not relativistic or mildly relativistic, $\Gamma\sim 1$ \citep{Izzo2019,Soker22} or relativistic $\Gamma\sim 100$ \citep{Fasano2021}. 
We estimate the neutrino flux for different sets of parameters.

Accelerated protons may lose energy due to synchrotron radiation and adiabatic losses. Following \cite{Fasano2021}, we compared the proton acceleration time with the photomeson loss time obtaining a cut-off energy value as:

\begin{equation}
    E_{\rm cut,15}^{\rm IS} \sim 0.066 \epsilon_{b,-1}^{1/2} \epsilon_{e,-1}^{-3/4} L_{50}^{5/8} \Gamma_2^{-3} t_{\rm jet,3}^{-1/4} \rho_{H-7}^{1/8}
\end{equation}

where $\epsilon_{b}$ is the fraction of energy converted in the magnetic field at the shock.  We assume equipartition implying  $\epsilon_{b}\sim 0.1$ 

%We obtained $E_{\rm p, \rm cut}=66$ TeV for $\Gamma=100$, $E_{p,cut}=66$ PeV for $\Gamma=10$. 

For energies $E_{\rm p}>E_{\rm p,\rm cut}$ the simple power law of Fermi acceleration is modified as follows:

\begin{equation}
    \frac{dN}{dE_p} \propto E_{\rm p}^{-2} e^{\frac{E_{\rm p}-E_{\rm p,cut}}{2E_{\rm p, cut}}} \qquad
    \end{equation}

The flux of emerging neutrinos can be suppressed due to pion and muon energy losses in the dense radiation field and the jet magnetic field. We take into account these effects. The proton synchrotron losses are the ones that most affect the neutrino spectra, this effect is stronger for larger Lorentz factor.

%Fixing the $L_{iso}$, the proton looses more energy with a larger $\Gamma$ factor, for $L_{iso}=10^{48}$ erg, we obtain a proton cutoff energy $E_{p,cut}\sim 10^{12}$ eV for $\Gamma=100$, and  $E_{p,cut}\sim 10^{18}$ eV for $\Gamma=1$. 

\subsection{Expected number of events}

\begin{figure}
    \centering
    \includegraphics[scale=0.45]{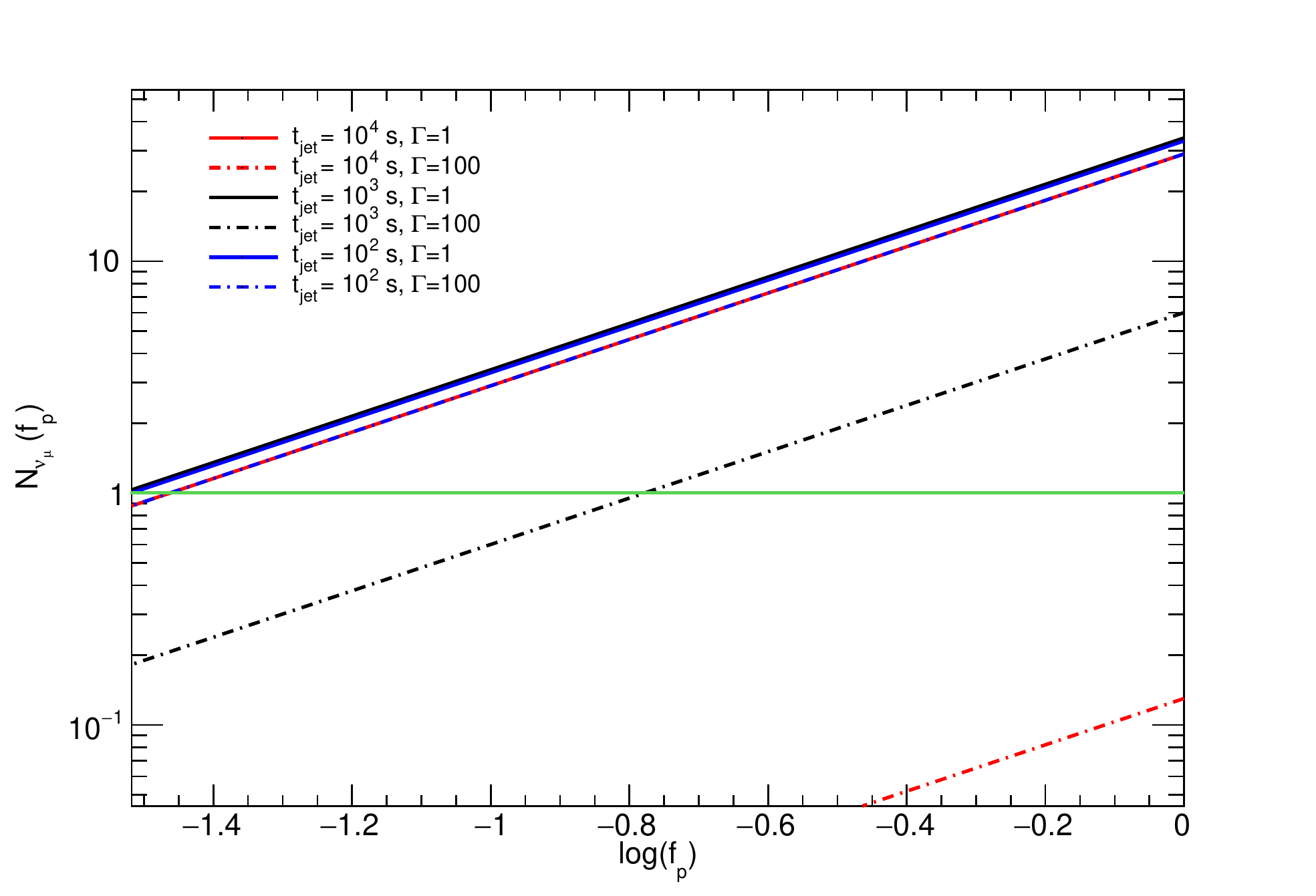}
    \caption{Kinetic energy of the jet $E=10^{51} erg$, the figure shows the muon and antimuon neutrino number in Icecube telescope, as a function of the baryon loading $f_p$ for different sets of the parameters $t_{jet}=10^2, 10^3, 10^4$ s and $\Gamma=1,100$. }
    \label{fig:10_51}
\end{figure}

Integrating the flux of neutrinos expected from this model convoluted with the detector area, we estimate the neutrino number of events expected from IceCube:

\begin{equation}
\label{Eq:Nevents}
    N_{\rm events}(\delta)= \int A_{\rm eff}^{\nu_\mu} (E_{\nu_{\mu}},\delta) \left( \frac{dN_{\nu_\mu}}{dE_{\nu_\mu}dS} \right)_{\rm Earth} dE_{\nu_\mu} \, 
\end{equation}

where $A_{eff}^{\nu_{\mu}}$ represents the effective area for IceCube \cite{aeffI}. We consider 
different hypotheses for the kinetic energy in the jet 
$E=10^{53}$ erg and 
$E=10^{51}$ erg, similar to the one considered by \cite{Murase2011}. Fig.\ref{fig:10_51} 
shows the neutrino number of events from SN2023ixf expected to be detected by IceCube as a function of $f_p$ for different sets of model parameters and $E=10^{51}$ erg. We can see from these figures that in the case of $f_p\sim1$ most of the models predict $>10$ number of neutrino events. Only for $\Gamma=100$ and $t_{\rm jet}=10^4$ sec the neutrino number of events is smaller than 1.

%\begin{figure}
 %   \centering
  %  \includegraphics[scale=0.6]{10_53.pdf}
   % \caption{Fixed $E_{iso}=10^{53} erg$, the figure shows the muon and antimuon neutrino number in Icecube telescope as a function of the baryon loading $f_p$ for different set of the parameters $t_{jet}=10^2, 10^3, 10^4$ s and $\Gamma=1,100$. }
   % \label{fig:10_53}
%\end{figure}

Another model for high-energy neutrinos considers a scenario in which the SN ejecta crashes into the dense CSM shells of the external medium \citep{Murase2011}. A significant fraction of the ejecta kinetic energy is converted to the internal energy of the shocked shells. Protons can be accelerated at the shocks and interact with the protons present in the external medium. Accelerated protons are mostly confined and produce mesons via inelastic pp scattering, which leads to the production of neutrinos. Within the \cite{Murase2011} model, assuming that all the energy goes into accelerating the protons, $\sim 50$ neutrinos with $E> 1$ TeV are expected at the distance of SN2023ixf.

\section{Observational predictions and implications}

In SN 1987A the neutrino burst was observed about three hours before the first (pre-discovery) electromagnetic detection. This time delay is consistent with the time taken for the SN shock break-out to emerge from the SN 1987A progenitor as a UV flash and mark the beginning of the SN explosion \citep{Arnett1987}. Therefore we would have expected to have detected a low-energy neutrino signal from SN2023ixf within a comparable time frame. We limited our search only to anti-electron neutrinos expected to be detected via Inverse-Beta Decay in both cases. Our analysis finds values that are below the detection threshold of the current Super-K and future Hyper-K neutrino telescope consistently with the lack of detected events confirmed by Super-K Collaboration since no excess signal was found in a time window of 2 days prior to the SN event \citep{atel_sk}.\\

The high-energy neutrino flux produced in the choked jet scenario should follow the SN photon emission. The delay time between photons and neutrinos, which is dominated by the time taken by the tail of the jet to catch the head, depends: i) on the size and location of the massive H envelope, ii) the duration of the jet, $t_{\rm jet}$, which can reach values up to $10^{4}s$  \citep{Stratta2015}; iii) the speed of the jet, $\Gamma \sim 1$ \citep{Izzo2019}. All of this implies a delay time of several hours.

Fig, 2 shows that, for a jet kinetic energy $\sim 10^{51}$ erg and $f_p\sim1$, then $\sim 30$ neutrinos are expected to be observed by IceCube for several jet parameters. However, \citep{IceCubeSN2023} were able to determine an upper limit for IceCube neutrino detections of less than 3 events. Thus, suggesting that, if the "jet scenario" is at play, the fraction of kinetic energy converted to accelerated protons is only $f_p<10\%$.

Considering the maximum number of events, $\sim 35$, obtained for $f_p\sim 1$, we can evaluate the distance $d_{\rm max}$ at which the source can be detected by IceCube. From the number of events at the distance $d_{L,1}= 6.4 $ Mpc, the new number reads as:

\begin{equation}
    N_{\rm events}(d_{\rm max})=   N_{\rm events}(d_{L,1}) \frac{d_{L,1}^2}{d_{\rm max}^2} 
\end{equation}

which implies:

\begin{equation}
    d_{\rm max}= \sqrt{\frac{N_{\rm events}(d_{L,1})}{{ N_{\rm events}(d_{\rm max})} } d_{L,1}^2 }
\end{equation}

For the set of parameters: $E=10^{51} \rm erg$, $f_p=1$, $\Gamma=1$, $t_{jet}=10^3 \rm s$ and $N_{\rm events}(d_{\rm max})\sim3$, we obtain a distance $d_{\rm max}\sim 22 $ Mpc. This distance roughly corresponds to the "Virgo circle" within which we expect to detect about one CC-SN per year, e.g. \citep{Cappellaro2015, Botticella2017,burderi2020} 
A similar analysis carried out on the effective area of KM3NeT produced estimates of neutrino detections comparable to those obtained for IceCube.
\smallskip

We can use also the IceCube results to provide some constraints on the \cite{Murase2011} "delayed" model. If the high-energy neutrinos are produced during the interaction with the dense circumburst material, the spectroscopic follow-up presented by \citep{Yamanaka2023} can help to set some interesting constraints. In the latter, authors show spectroscopic signatures of the interactions of the SN ejecta with the CSM lasting for less than 2 weeks from maximum light. Since the SN ejecta moves at a speed that is significantly smaller than that of the jet cocoon (e.g. Fig. 3 in \cite{Izzo2019}), one could infer that a burst of "delayed" high energy neutrinos might be observed within $\sim 2$ weeks after the peak of light, or so. IceCube did not detect any neutrino from this source till now \citep{IceCubeSN2023} and this fact implies that also in the Murase model, the fraction of energy that goes in accelerated protons should be smaller than $10\%$.

The expected number of neutrinos from nearby galaxies has been explored in many other studies \citep{Murase2018,Murase2022,Murase2023,valtonen2023}, that consider a model similar to the one described in \cite{Murase2011}. In this paper, we have considered a model where the jet parameters like its Lorentz factor and duration can be constrained by the observations.

\cite{Murase2018} shows that both the expected neutrino number and the
optimal time window for neutrino observations depend on the type of
CCSN. In our analysis, we considered SN2023ixf to be a type II-P and assumed a time window ($<10^4$ sec) after the SN alert.

%\cite{Murase2018} shows that the expected number of neutrinos and the optimal time for observation in neutrino telescopes depend on the type of CCSN. In our analysis, we have considered a \textbf{type II-P} SN and 
%have assumed a short time window ($<10^4$ sec) after the detection of SN2023ixf in agreement with the times suggested in \cite{Murase2018}. 

%\textbf{ We assume that the total duration for calculating the neutrino signal events is $\sim 10^3-10^4 $ sec}.

Applying the upper limits found by IceCube to this source \citep{IceCubeSN2023},
we find similar results to previous estimates on the fraction of accelerated protons $<10\%$.

\bigskip

\section{Conclusions} 
\bigskip

Some interesting results emerge from our analysis:
\bigskip

i) Fig.\ref{fig:sk_events} shows that low-energy neutrinos produced in a Galactic SN explosion (e.g. Betelgeuse or in the Galactic Center) can be easily detected by Super- and Hyper-Kamiokande observatories providing several $\times 10^{5-6}$ events. The detection of MeV neutrinos is also within the capabilities of Super-K for SN explosions occurring in the LMC and M31. Detections outside the Local Group of Galaxies appear more problematic, as shown by the recent explosions of SN 2023ixf at 6.4 Mpc. For this SN, no neutrinos have been observed by the Super-K Collaboration in \cite{atel_sk}, and we estimate $\sim 1$ event even from Hyper-K. In other words, the bottleneck to future low-energy neutrinos detections is represented by the rate of core-collapse events expected for our Galaxy, about one event every 60 years \citep{Cappellaro2015,Rozwadowska2021} and about 1 CC-SN in $\sim 30$ years within the Local Group of Galaxies. 

ii) A close inspection of Fig.\ref{fig:10_51} finds that the lack of high-energy neutrinos detection from IceCube \citep{IceCubeSN2023} can be explained in different ways:
a) only a small fraction of the kinetic energy, $f_p<10\%$ is converted in accelerating protons at the shocks;
b) if the duration of the jet is very long $t_{\rm jet}\sim 10^{4}$ sec and the Lorentz factor is rather high, i.e. $\Gamma\sim 100$, the protons accelerated at the shock will lose all their energy in synchrotron emission and will not interact with photons to produce neutrinos. In this case, $f_p$ can be $\sim 1$; c) the jets are not driving the explosion of SN 2023ixf. On the other hand, this SN appears to be a relatively "standard" type II core collapse event, both in terms of size and mass of the progenitor \citep{Kilpatrick2023}, and spectroscopic \citep{Yamanaka2023} and photometric evolution \citep{Hosseinzadeh2023}. 
This conclusion is consistent with the results reported by \cite{icecube_2023} who found no significant spatial or temporal correlation of neutrinos with cataloged supernovae. In fact, their analysis found that the IIn and IIP SNe emit less than $1.3 \times 10^{49}$ erg and $2.4 \times 10^{48}$ erg in high-energy neutrinos.
The lack of detection of high-energy neutrinos might suggest that the jet scenario might be distinctive of stripped-envelope SNe (i.e. Ic types) associated with long-duration GRBs, rather than being a mechanism common to all collapse events.
Obviously, this point will be clarified by future observations of nearby SNe.

iii) We focused our analysis assuming $\Gamma \sim 1$ and a kinetic energy of the jet E$\sim 10^{51}$erg because they are values  directly suggested by observations, \citep{Izzo2019,Filippenko1997}. We have also explored more extreme cases $\Gamma\sim 100$ and E$\sim 10^{53}$erg and obtained more stringent constraints on baryon loading, $f_p < 1\%$. Given an average rate of about one core-collapse SN per year, within the "Virgo circle", the latter assumptions may be tested within a few years from now.
\bigskip
\bigskip

%\acknowledgements
\section*{acknowledgements}
AL and MDV are grateful to Ariel University for the hospitality and pleasant atmosphere. The authors thank the referee for her/his constructive criticisms that
have improved the presentation and discussion of the data.

\clearpage
\bibliographystyle{aasjournal}
\bibliography{biblio}

%\printbibliography

\end{document}